\documentclass[letterpaper,10pt]{article} 

\usepackage{opticameet3} 

\newcommand\authormark[1]{\textsuperscript{#1}}

\usepackage{amsmath,amssymb}
\usepackage[colorlinks=true,bookmarks=false,citecolor=blue,urlcolor=blue]{hyperref} 

\begin{document}

\title{Long-term cybersecurity applications enabled by quantum networks}

\vspace{-0.25in}
\author{Nicholas A. Peters,\authormark{1,*} Muneer Alshowkan,\authormark{1} Joseph C. Chapman,\authormark{1} Raphael C. Pooser,\authormark{1}  Nageswara S. V. Rao,\authormark{2} and Raymond T. Newell \authormark{3}}

\address{\authormark{1} Quantum Information Science Section, Oak Ridge National Laboratory, Oak Ridge, TN 37831\\
\authormark{2}Advanced Computing Methods for Engineered Systems Section, Oak Ridge National Laboratory, Oak Ridge, TN 37831\\
\authormark{3}MPA-Quantum, Los Alamos National Laboratory, Los Alamos, NM 87545}

\email{\authormark{*}petersna@ornl.gov} 

\vspace{-0.25in}

\begin{abstract}
%
If continental-scale quantum networks are realized, they will provide the resources needed to fulfill the potential for dramatic advances in cybersecurity through quantum-enabled cryptography applications.  We describe recent progress and where the US is headed as well as argue that we go one step further and jointly develop quantum and conventional cryptography methods for joint deployments along the quantum backbone infrastructure.

\end{abstract}

\vspace{0.25in}

\textit{Challenge.}
Fault-tolerant universal quantum computing has the potential to make the majority of current cryptographic protocols obsolete.  This is colloquially known as ``the quantum computing threat.'' As a result, crucial cryptographic functions that provide confidentiality, integrity, and availability of the communications which underpin global infrastructures are potentially at risk. This risk extends to the classical communications systems required to operate a quantum computer or quantum network; control signals and human-readable data are classical, and are therefore just as vulnerable as any other IT system to eavesdropping, spoofing, or other cyber attack.  As we develop and deploy quantum networks, it will be essential to include security in their design from the beginning, rather than as a retrofit at the end. 

There are two lines of defense against a cryptography-breaking quantum computer: classical cryptography systems which derive their strength from math problems which remain hard even with a quantum computer (often called Post-Quantum Cryptography, PQC); and quantum cryptographic systems (QCS) which derive their strength from the fundamental laws of physics. The first category, PQC, is under active development worldwide, including an ongoing competition sponsored by NIST to select and standardize quantum-safe cryptosystems~\cite{NIST2016}. On the other hand, QCS require development of separate hardware for their deployment; examples include quantum key distribution (QKD), which is the most mature QCS protocol, quantum digital signatures (QDS), and quantum secret sharing (QSS).

As a result of new hardware development for QCS, there are multiple research challenges that need to be addressed for QCS to realize its full potential. As QCS is a hardware-based solution, it is currently very expensive.  For example, most discrete-variable QCS systems (e.g., encoding in polarization or time bin) utilize direct single photon detection (DD) with costly single-photon detectors (at least in the context of most modern telecommunications fiber networks). Additionally, DD-QCS can be severely limited by Raman scattering of classical light used to carry data~\cite{peters2009}. Continuous variable (CV) approaches (encoding in amplitude and phase) utilize homodyne detection which is more cost effective, relatively immune to Raman scattering~\cite{Qi_2010}, and highly efficient during room-temperature operation. As a result, integration of DD-QCS into optical networks is challenging without very strict limitations on conventional data signals carried in the same fiber. In contrast, CV-QKD can be deployed with multiple optical channels carrying commercial levels of data. However, the DD-QCS is much more mature than CV-QCS, and for example, additional assumptions are frequently made for CV-QKD security about the detection process. In either case, most QCS systems are still expensive laboratory experiments or in bulky rack-mounted boxes with limited ruggedness for deployment. Moreover, it is an important open research question as to how to best securely implement and certify QCS, for example, so that side channels do not leak unintended information.

In addition, QCS assumes that an authenticated classical communications channel is available for the after-quantum transmission processing of the protocol.  Much of the current cryptography infrastructure is public (asymmetric) key based whereas QKD delivers (private) symmetric keys, so QKD is not a drop-in replacement of current infrastructure.  While QDS and QSS are multi-party protocols, they are not drop in replacements for existing cryptography methods either. As a result, how to authenticate the classical conventional channel and how to best utilize QCS in existing infrastructures remain a research challenge.

\textit{Opportunity.}
PQC is conventional cryptography thought to be secure against significant quantum computers and is hoped to be secure against foreseeable technology developments for at least several decades.  On the other hand, QCS could potentially enable security for much longer time scales since the security is dependent on physics, which is technology agnostic, instead of on computational difficulty like PQC. 
This is an attractive feature for securing critical infrastructure, which could include science networks as well as energy systems, which are often expected to last for decades or longer.

Furthermore, QCS networks can benefit from using satellite-based quantum networks because satellites are difficult to access and can be monitored to ensure they remain physically secure, at least as secure as any ground station and probably more secure. The trade-off is that ground stations are required to communicate with the satellite; but, rugged portable ground stations are being commercialized. Satellite links could also enable longer-distance QCS links sooner than waiting for quantum repeaters to be developed~\cite{Zhang:18}, even benefiting from satellites in geostationary orbit that allow for more continuous key generation~\cite{PhysRevApplied.18.044027}. As a concrete example, the development of small rugged high-performance QKD satellites serving as ``trusted nodes'' would provide almost immediate practical benefit by distributing usable cryptographic keys to interested users on global scale.

Moreover, measurement-device-independent and device-independent implementations~\cite{PhysRevLett.108.130503,PhysRevLett.98.230501} significantly reduce the security  requirements and assumptions.  These more advanced protocols are designed to be secure even if certain parts of the hardware are not able to be located in a secure location. There have been demonstrations of measurement-device-independent QKD.  Given the further increased security of these measurement-device-independent implementations over standard QCS (which already has advantages over other cryptography protocols relying on assumed computational difficulty) they could enable more flexibility in how QCS systems are deployed.

\textit{Assessment.} 
In contrast to much of the rest of the world, in recent years, QCS research has not been a major focus in the US, where the focus is on PQC.  The National Institute of Standards and Technology (NIST) is working to standardize PQC to counter the quantum computing threat \cite{NIST2016}.  It does not seem possible to standardize QCS and related technologies, using the same process as PQC as they are rooted in fundamentally different ideas.  In addition, QCS is relatively immature compared to conventional cryptography. Nevertheless, for QKD in particular, there are natural metrics of secure key rate which are common to compare performance between different implementations. Additionally, device-independent and measurement-device-independent QCS protocols provide security verification that can be certified via loop-hole free Bell tests and Bell state measurements, respectively~\cite{PhysRevLett.108.130503,PhysRevLett.98.230501,Bierhorst2018}.

\textit{Timeliness or maturity.}
Despite the research challenges, the promise of long-term security independent of computational capability has caused QCS to be the focus of numerous academic research and corporate development programs globally.  And while it is relatively immature compared to conventional cryptography, it is already a commercialized nearer-term application enabled by quantum networks. Even though there are several commercial offerings, much research and development must be done to close the gaps we describe. This research would hopefully make products more secure, enable their certification, enable longer communications distances, and increase their secure key rates. When fully mature, QCS protocols could enable long-term security of future quantum networking infrastructures.

\bibliographystyle{opticajnl.bst}
\bibliography{QKDwhitepaper}

\end{document}